\documentclass[conference]{IEEEtran}
\usepackage{graphicx}
\usepackage[center]{caption}
\usepackage{subcaption}
\usepackage{cite}
\usepackage{color}
\usepackage{amsmath,amssymb,amsthm}

\newcommand\norm[1]{\left\lVert#1\right\rVert}
\usepackage{algorithm,algorithmic,multicol}

\begin{document}
\title{Flexible Hybrid Beamforming for Spectrally Efficient 6G Joint Radar-Communications}

\author{\IEEEauthorblockN{Aryan Kaushik$^1$, Evangelos Vlachos$^2$, Muhammad Z. Shakir$^3$, Wonjae Shin$^4$, Rongke Liu$^5$}
\IEEEauthorblockA{
$^1$Department of Engineering and Design, University of Sussex, UK \\
$^2$Industrial Systems Institute (ISI), Athena R\&I Center, Greece\\
$^3$School of Computing, Engineering \& Physical Sciences, University of the West of Scotland, UK \\
$^4$Department of AI Convergence Network, Ajou University, South Korea\\
$^5$School of Electronics \& Information Engineering, Shenzhen Institute of Beihang University, China\\
Emails: aryan.kaushik@sussex.ac.uk, evlachos@athenarc.gr, muhammad.shakir@uws.ac.uk, \\wjshin@ajou.ac.kr, rongke\_liu@buaa.edu.cn}}

\maketitle

\begin{abstract}
Joint radar-communications (JRC) benefits from multi-functionality of radar and communication operations using same hardware and radio frequency (RF) spectrum resources. Thus JRC systems possess very high potential to be employed into the sixth generation (6G) standards. This paper designs a flexible beamformer for multiple-input multiple output (MIMO) JRC with maximized spectral efficiency (SE). Hybrid beamforming is implemented which constitutes lesser number of RF chains than number of transmitter antennas. We jointly express JRC rate with communication and radar entities including a weighting factor which depicts the dominance of one operation over the other. The joint-SE based proposed method optimally selects the number of RF chains with flexible hynrid beamforming design. Furthermore, when the communication operation takes place the proposed method takes into account the interference occurring from the radar operation and vice-versa. Fractional programming based selection procedure is used for flexible beamforming and optimal number of RF chains while considering interference of each operation. Simulation results are presented and compared with different baselines to show effectiveness of the proposed flexible hybrid beamforming method.
\end{abstract}

\begin{IEEEkeywords}
Joint radar-communications, flexible beamforming, RF selection, interference, hybrid precoder. 
\end{IEEEkeywords}

%
\IEEEpeerreviewmaketitle

\section{Introduction}
Mobile connectivity is expected to be over 70\% of the global population, and fifth generation (5G) connections are estimated to constitute towards 1.4 billion mobile devices by 2023 \cite{cisco2020}. By 2027, 5G subscriptions will be 4.4 billion presenting faster growth than previous generation standards, and average monthly usage per smartphone reaching to 40 GB globally, and 52 GB of data traffic in both Western Europe and North America \cite{ericsson2022}. Thus to initiate the implementation of sixth generation (6G) wireless standards, more advanced technical solutions such as sharing of hardware and spectral resources for several wireless applications, are required to make efficient use of the available resources. This will lead to decongestion of existing sub 6-GHz spectrum which implements majority of the current mobile services, and solve hardware inefficiency issues with massive connectivity. 

One such approach is to unify sensing and communications operations on a single hardware and use of same spectral resources, where sensing collects and extracts information, such as target detection, 
and using radio waves for tracking movement, while communication possesses transfer of information \cite{paulAccess2017, zhangJSTSP2021, liuTCOM2020}. The emerging joint radar and communications (JRC) systems accommodate multiple functionalities of these operations, massive connectivity with limited resources, and provide wide applications in future wireless, defence, aerial, vehicular, internet-of-things and space \cite{zhangJSTSP2021, liuTCOM2020, cuiNet2021, aryanIET2022}.
To achieve high beamforming gains with high degrees of freedom, JRC can employ multiple-input multiple-output (MIMO) antennas which improves range resolution and also makes up for the high path loss associated with high-frequency bands such as millimeter wave \cite{liuTWC2018, mishraSP2019}.

Hybrid precoding architecture, which constitutes of analog and digital precoding, has been implemented for 5G MIMO systems to obtain energy and hardware efficient designs such as in \cite{aryanTGCN2019, aryanTGCN2021, vlachosRsoc2020}. Recently, hybrid precoding has been implemented in energy and hardware efficient JRC \cite{aryanICC2021} and it is also implemented with low resolution sampling based digital-to-analog converters (DACs) in JRC \cite{aryanICC2022}. Advanced multiple access approaches such as the non-orthogonal multiple access (NOMA) and rate splitting multiple access (RSMA) have also been incorporated with JRC \cite{wangCL2022, dizdarOJCOMS2022} for interference management and massive connectivity. Further advanced signal processing approaches in JRC implement low complexity waveform designs and index modulation based JRC with antenna and frequency agility \cite{aryanJCNS2022, shenGCOM2022, huangTCOM2020}. However, latest JRC systems do not consider jointly designing rate for both the operations with flexible beamforming while taking into account the interference occurring from each of the operations in dual function JRC. 

\subsubsection*{Contributions} 
In this paper, for 6G JRC systems, we jointly design radar-communication rate in terms of weighting formulation which decides the dominance of one operation over the other while both radar and communication operations take place simultaneously. We employ this formulation to solve an interference-oriented spectral efficiency (SE) maximization problem while interference minimization takes place and power constraints are imposed on both the operations. We develop a selection procedure based on fractional programming which is employed to optimize the number of available RF chains in dual function JRC systems while interference is minimized and joint-SE is maximized. This procedure leads to highly spectral efficient design with low hardware complexity (via RF chain selection). Such framework carries high potential to be used in the future 6G wireless communication standards. Numerical results show communication and radar performance gains of joint SE-based proposed approach over baselines for different weighting factor values and interference scenarios.

\emph{Notation:} 
The notation \textbf{M} represents matrix, \textbf{m} represents vector, m is scalar entity, tr(.) and $|.|$ represent trace and determinant functions,, $(.)^{T}$ denotes transpose, $(.)^{H}$, $\norm{.}_F$ and $\Vert.\Vert_2$ represent complex conjugate transpose, Frobenius norm, and L2 norm, respectively. The notation $[\textbf{M}]_{m,k}$ is $(m,k)$-th element in matrix $\mathbf{M}$ and $\mathbf{m}_m$ is $m$-th element of $\mathbf{m}$; $\textbf{I}_{J}$ is $J$-size identity matrix; $\mathbf{I}_{J \times K}$ denotes column concatenated matrix as $[\mathbf{I}_J \,\, \mathbf{0}_{J \times K}]$.
$\mathbb{C}$ and $\mathbb{R}$ represent sets of complex and real numbers, respectively, and $\mathbb{E}$ denotes expectation operator. The notation $\mathcal{C}\mathcal{N} (m, n)$ denotes complex Gaussian vector which has mean $m$ and variance $n$. 

\section{System Model}
The JRC system with MIMO antenna setup has $N_\textrm{T}$ number of antennas at the transmitter, $N_\textrm{R}$ represents number of antennas at the mobile users (UEs), and for radar sensing, number of targets are represented by $N_\textrm{p}$. A hybrid beamforming architecture is considered which results into fewer number of available RF chains than antennas, represented by $L_\textrm{T}$ and $1\leq L_\textrm{T}\leq N_\textrm{T}$. $\mathbf{s} \in \mathbb{C}^{N_\textrm{R} \times 1}$ represents the transmit signal where $\mathbb{E}\{\mathbf{s} \mathbf{s}^H\} = \mathbf{I}_{N_\textrm{R}}$. The transmit signal constitutes $\mathbf{s} = \left[ \begin{array}{cc} \mathbf{s}_\text{com}^T & \mathbf{s}_\text{rad}^T \end{array}\right]^T$, where $\mathbf{s}_\text{com} \in \mathbb{C}^{\frac{N_\textrm{R}}{2} \times 1}$ refers to the communication vector term and $\mathbf{s}_\text{rad} \in \mathbb{C}^{\frac{N_\textrm{R}}{2} \times 1}$ is for the radar operation. Furthermore, dual function transmission corresponds to $\norm{\mathbf{s}_\text{com}}^2 = \norm{\mathbf{s}_\text{rad}}^2 =1$. 

As shown in the block diagram of Fig. 1, hybrid precoding is implemented at the transmitter for JRC where parameters for both the communication and radar operations are represented with different blocks. 
In Fig. 1, digital precoder is represented by $\textbf{F}_\textrm{BB}$ followed by the RF Chains. The analog precoder is represented by $\mathbf{F}_\textrm{RF}$ which consists of a network of phase shifters. The elements of analog precoder with constant-modulus entries are denoted as $f_\textrm{i} \in \mathbb{C}^{\frac{N_\textrm{T}}{L_\textrm{T}}\times 1}, \forall i = 1,.., L_\textrm{T}$. The decomposition $\mathbf{F}_\textrm{BB}^\text{com}\mathbf{F}_\textrm{RF}^\text{com}$ is used for the digital and analog precoding when communication operation takes place, and $\mathbf{F}_\textrm{BB}^\text{rad}\mathbf{F}_\textrm{RF}^\text{rad}$ accounts for digital and analog precoding when radar sensing operation takes place. The RF selection procedure is implemented at the baseband unit in the JRC system.

    \begin{figure}
 \begin{center}
 	\includegraphics[width=0.75\textwidth, trim=0 100 40 140,clip]{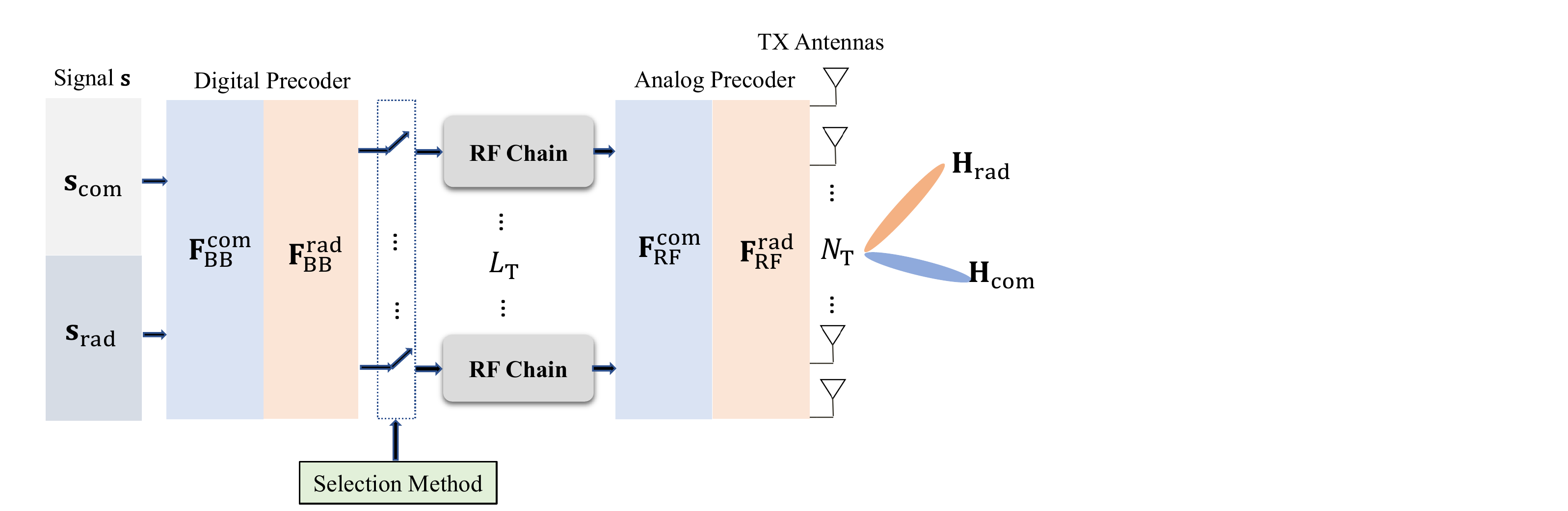}
 \end{center}
 	\caption{Flexible hybrid beamforming based JRC system.}
\end{figure}

The JRC transmit signal $\mathbf{x}_\textrm{T}$ can be written as
\begin{align}
\mathbf{x}_\textrm{T} 
&= \mathbf{F}_\textrm{RF}^\text{com}\mathbf{F}_\textrm{BB}^\text{com} \mathbf{s}_\text{com} + \mathbf{F}_\textrm{RF}^\text{rad}\mathbf{F}_\textrm{BB}^\text{rad} \mathbf{s}_\text{rad}\nonumber \\
&+\underbrace{\mathbf{F}_\textrm{RF}^\text{com}\mathbf{F}_\textrm{BB}^\text{rad} \mathbf{s}_\text{rad} + \mathbf{F}_\textrm{RF}^\text{rad}\mathbf{F}_\textrm{BB}^\text{com} \mathbf{s}_\text{com}}_{\text{interference term}},
\end{align}
where in the transmitting unit, interference term represents the interference from radar operation to the communication and communication operation to the radar.

We consider narrowband channel model for communication which is mostly used for millimeter wave channel, however our proposed approach is independent of the channel modeling, and for instance, it can be applied throughout from existing sub 6-GHz bands to terahertz (THz) frequency spectrum. The communication channel is represented as $\mathbf{H}_\text{com} \in \mathbb{C}^{N_\textrm{R}\times N_\textrm{T}}$ and expressed as
\begin{equation}\label{eq:channel_model}
\mathbf{H}_\text{com} = \sqrt{\frac{N_\textrm{T}N_\textrm{R}}{N_\textrm{c}}} \sum_{l=1}^{N_{\textrm{c}}}  \alpha_{l} \mathbf{a}_{\textrm{R}}(\phi_{l}^{r}) \mathbf{a}_{\textrm{T}}^H(\phi_{l}^{t}),
\end{equation}
with $\alpha_{l}$ being the channel gain of $l$-th path, and $N_\textrm{c}$ being the number of clustered multipaths. The term $\textbf{a}_{\textrm{T}}(\phi_{l}^{t}) = \frac{1}{\sqrt{N_{T}}}{[1, e^{j \frac{2 \pi}{\lambda}d\sin(\phi_{l}^{t})}, ..., e^{j (N_{T}-1)\frac{2 \pi}{\lambda}d\sin(\phi_{l}^{t})}]}^{T}$ represents the steering vector for transmission. The departure angle is represented as $\phi_{l}^{t}$ with $d$ being spacing between antennas and $\lambda$ being wavelength. Similarly, the term $\textbf{a}_{\textrm{R}}(\phi_{l}^{r})$ represents array response vector for receiver. The angle of arrival is represented as $\phi_{l}^{r}$. It is assumed that the channel state information of communication channel is known at transmitter and UE sides. Furthermore, the antenna setup in \eqref{eq:channel_model} follows an uniform linear array (ULA) configuration while the proposed method is independent of antenna configuration, for instance, it can also be applied to circular array and rectangular array configurations.  The unknown radar channel is represented as $\mathbf{H}_\text{rad}$, where covariance matrix $\mathbf{R}_\textrm{T}$ will be dependent on the columns of radar channel and is considered to be a known parameter. 

For radar sensing operation, in terms of the beampattern at transmitter which is pointed towards the targets of interest, we can express the following:
\begin{equation}\label{eq:beampattern}
    P_\textrm{T}(\phi_i) = \mathbf{a}^H_\textrm{T}(\phi_i) \mathbf{R}_\textrm{T} \mathbf{a}_\textrm{T}(\phi_i),
\end{equation}
where we consider that targets, i.e, $N_\textrm{p}$, are placed at different angles $\phi_i$, with $i=1,..,N_\textrm{p}$. 
The design of transmit beampattern $P_\textrm{T}(\phi_i)$, in \eqref{eq:beampattern}, is equivalent to designing the parameter $\mathbf{R}_\textrm{T}$ which can be represented in terms of the hybrid precoder decomposition, i.e., $\mathbf{R}_\textrm{T} = \mathbf{F}_{\textrm{RF}} \mathbf{F}_{\textrm{BB}} (\mathbf{F}_{\textrm{RF}} \mathbf{F}_{\textrm{BB}})^H$. 
The optimal radar precoder $\mathbf{F}_\text{rad}^\textrm{opt}$ (with diagonal entries) consists of $\mathbf{v}_\textrm{i} \in \mathbb{C}^{\frac{N_\textrm{T}}{N_\textrm{p}}\times 1}$ elements composed by $\mathbf{a}_\textrm{T}(\phi_i)\,\forall i = 1,.., N_\textrm{p}$ entries, located at corresponding slots. The radar covariance matrix $\mathbf{R}_\textrm{T}^\textrm{opt}$ related to the optimal radar precoder $\mathbf{F}_\text{rad}^\textrm{opt}$ is given by, $\mathbf{R}_\textrm{T}^\textrm{opt} = \mathbf{F}_\text{rad}^\textrm{opt}(\mathbf{F}_\text{rad}^\textrm{opt})^H$ corresponding to a well-designed radar beampattern case. Next, we will proceed with the problem formulation and flexible beamforming based propose method for RF selection.

\section{SE Maximization for JRC}\label{se_maximization}
This section first describes the weighted SE formulation (with a weighting factor) for radar and communication operations which also takes into account interference terms, and further we address the SE maximization optimization problem. Afterwards, we propose a \textit{flexible beamforming} design, by selecting the optimal subset of RF chains which will be activated using the proposed selection method.

\subsection{Problem Formulation}
First we express the communication rate as follows:
\begin{align}\label{eq:rate_comms}
&R_\text{com}(\mathbf{F}_\textrm{RF}^\text{com}, \mathbf{F}_\textrm{BB}^\text{com})  = \log_2 \bigg(1 + \frac{1}{\sigma_{\text{rad-com}}^2} \times \nonumber \\ &\mathbf{w}_\text{com}^H\mathbf{H}_\text{com}\mathbf{F}_\textrm{RF}^\text{com} \mathbf{F}_\textrm{BB}^\text{com}{\mathbf{F}_\textrm{BB}^\text{com}}^H{\mathbf{F}_\textrm{RF}^\text{com}}^H \mathbf{H}_\text{com}^H\mathbf{w}_\text{com}\bigg),
\end{align}
and
\begin{equation}
\sigma_\textrm{rad-com}^2 \triangleq \mathbf{w}_\text{rad}^H\mathbf{H}_\text{rad}\mathbf{F}_\textrm{RF}^\text{com}\mathbf{F}_\textrm{BB}^\text{com}{\mathbf{F}_\textrm{BB}^\text{com}}^H{\mathbf{F}_\textrm{RF}^\text{com}}^H \mathbf{H}_\text{rad}^H\mathbf{w}_\text{rad},
\end{equation}
where $\sigma_\textrm{rad-com}^2$ corresponds to the radar interference while communication operation is taking place in the dual function JRC system.

Similarly, following \cite{aryanIET2022} and literature within, we can express the radar rate as follows:
\begin{align}\label{eq:rate_radar}
&R_\text{rad}(\mathbf{F}_\textrm{RF}^\text{rad}, \mathbf{F}_\textrm{BB}^\text{rad})  = \log_2 \bigg(1 + \frac{1}{\sigma_{\text{com-rad}}^2} \times \nonumber \\ &\mathbf{w}_\text{rad}^H\mathbf{H}_\text{rad}\mathbf{F}_\textrm{RF}^\text{rad} \mathbf{F}_\textrm{BB}^\text{rad}{\mathbf{F}_\textrm{BB}^\text{rad}}^H{\mathbf{F}_\textrm{RF}^\text{rad}}^H \mathbf{H}_\text{rad}^H\mathbf{w}_\text{rad}\bigg).
\end{align}
and
\begin{equation}
\sigma_\textrm{com-rad}^2 \triangleq \mathbf{w}_\text{com}^H\mathbf{H}_\text{com}\mathbf{F}_\textrm{RF}^\text{rad}\mathbf{F}_\textrm{BB}^\text{rad}{\mathbf{F}_\textrm{BB}^\text{rad}}^H{\mathbf{F}_\textrm{RF}^\text{rad}}^H \mathbf{H}_\text{com}^H\mathbf{w}_\text{com},
\end{equation}
where $\sigma_\textrm{com-rad}^2$ corresponds to the communication interference while the radar operation takes place in dual function JRC system.

Thus, following \eqref{eq:rate_comms} and \eqref{eq:rate_radar}, the joint radar-communication rate can be written into the following form:
\begin{equation}
    R = \rho R_\text{com} + (1-\rho)R_\text{rad},
\end{equation}
with $\rho$ being the weighting term between radar and communication operations which describes the dominance of one operation over the other, depending on its value between $0$ and $1$,  i.e., $\rho \in [0,1]$. For instance, when $\rho$ value is high, JRC system prioritizes communication operation and when $\rho$ value is low, JRC system prioritizes radar sensing operation.

Next, for the joint case in JRC considering the weighting allocated to each operation, SE maximization can be expressed in the following form: 
\begin{align}\label{eq:max_joint_rate}
\max_{\mathbf{F}_\textrm{RF}^\text{com}\mathbf{F}_\textrm{BB}^\text{com}, \mathbf{F}_\textrm{RF}^\text{rad}\mathbf{F}_\textrm{BB}^\text{rad}} & \rho R_\text{com} + (1-\rho)R_\text{rad} \nonumber \\
\textrm{ subject to } & \textrm{tr}(\mathbf{F}_\textrm{RF}^\text{com}\mathbf{F}_\textrm{BB}^\text{com} {\mathbf{F}_\textrm{BB}^\text{com}}^H{\mathbf{F}_\textrm{RF}^\text{com}}^H) \le P_\textrm{max}^\text{com} \nonumber \\
& \textrm{tr}(\mathbf{F}_\textrm{RF}^\text{rad}\mathbf{F}_\textrm{BB}^\text{rad} {\mathbf{F}_\textrm{BB}^\text{rad}}^H{\mathbf{F}_\textrm{RF}^\text{rad}}^H) \le P_\textrm{max}^\text{rad} 
\end{align}
where the terms $P_\textrm{max}^\text{com}$ and $P_\textrm{max}^\text{rad}$ represent the maximum power budget for communication and radar operations, respectively.

\begin{algorithm}[t]
	\caption{Proposed Method for RF-Selection in JRC}
	\label{algorithm:proposed}
	\begin{algorithmic}[1]
	    \REQUIRE $\mathbf{H}_\text{com}$, $\mathbf{H}_\text{rad}$
	    \ENSURE Diagonal binary matrices $\mathbf{S}^{\text{com}}$ and $\mathbf{S}^{\text{rad}}$
	    \STATE Initialize $\kappa^{(0)} = 1$.
		\FOR {$i=1,2,\ldots, I_{\rm max}$}
		\STATE Compute the term $\delta_{\textrm{com}}$ in \eqref{eq:delta_com}.
		\STATE Compute the term $\delta_{\textrm{rad}}$ in \eqref{eq:delta_rad}.
        \STATE Calculate the interference term $\sigma^2_{\textrm{rad-com}}(\mathbf{S}^{\text{com}})$ in \eqref{eq:sigma_rad_com}.
  	\STATE Calculate the interference term $\sigma^2_{\textrm{com-rad}}(\mathbf{S}^{\text{rad}})$ in \eqref{eq:sigma_com_rad}.
		\STATE Use CVX \cite{cvx} to solve \eqref{eq:S_problem_com}.
		\STATE Compute $\kappa_{(i)}^{\text{com}}$ and $\kappa_{(i)}^{\text{rad}}$ from \eqref{eq:kappa_computation}.  
		\ENDFOR
	\end{algorithmic}
\end{algorithm}

\subsection{Flexible Hybrid Beamforming}
Next, the joint SE maximization problem can be considered in the form of sparse subset selection based problem. This can be executed by introducing a sparse RF--chain selection diagonal matrix. For instance, $\mathbf{S}^\text{com}$ represents the diagonal selection matrix when communication operation takes place. The selection matrix consists of diagonal entries from the set with values $\{0,1\}$. It can be inferred that the diagonal matrix $\mathbf{S}^{com} \in \{0, 1\}^{L_\textrm{T} \times L_\textrm{T}}$ has binary diagonal entries and represents the activation or deactivation of the RF chains based on these entries, i.e., $[\mathbf{S}^{com}]_{ii} \in \{0, 1\}$, and also $[\mathbf{S}^{com}]_{ij}=0$ for $i \neq j \, \forall i=1,.., L_\textrm{T}$. Furthermore, hybrid precoder decomposition for communication operation with $\mathbf{S}^\text{com}$ selection matrix can be written as following:
\begin{equation}
    \mathbf{F}^\text{com} = \mathbf{F}_\textrm{RF}^\text{com} \mathbf{S}^\text{com} \mathbf{F}_\textrm{BB}^\text{com}.
\end{equation}
Similarly we can represent the diagonal selection matrix for radar sensing as $\mathbf{S}^\text{rad}$ for RF chain selection procedure when radar operation takes place. Note that, for brevity in the following, we do not use super index for the selection matrix, and express only $\mathbf{S}$ matrix term for both radar sensing and communication operations. 

Furthermore, we can express the communication rate as:
\begin{align}
    R_\text{com}(\mathbf{S}^{\text{com}}) =& \log_2 \Bigg(1 + \frac{1}{\sigma_\textrm{rad-com}^2} \mathbf{w}_\text{com}^H \mathbf{H}_\text{com} \mathbf{F}_\textrm{RF}^\text{com} \mathbf{S}^{\text{com}} \mathbf{F}_\textrm{BB}^\text{com} \times \Bigg. \nonumber \\ &\hspace{9mm}\Bigg.{\mathbf{F}_\textrm{BB}^\text{com}}^H \mathbf{S}^{\text{com}} {\mathbf{F}_\textrm{RF}^\text{com}}^H \mathbf{H}_\text{com}^H \mathbf{w}_\text{com}\Bigg),
\end{align}
and following that, radar rate $R_\text{rad}(\mathbf{S}^{\text{rad}})$ can be written as
\begin{align}
    R_\text{rad}(\mathbf{S}^{\text{rad}}) =& \log_2 \Bigg(1 + \frac{1}{\sigma_\textrm{com-rad}^2} \mathbf{w}_\text{rad}^H \mathbf{H}_\text{rad} \mathbf{F}_\textrm{RF}^\text{rad} \mathbf{S}^{\text{rad}} \mathbf{F}_\textrm{BB}^\text{rad} \times \Bigg. \nonumber \\ &\hspace{9mm}\Bigg.{\mathbf{F}_\textrm{BB}^\text{rad}}^H \mathbf{S}^{\text{rad}} {\mathbf{F}_\textrm{RF}^\text{rad}}^H \mathbf{H}_\text{rad}^H \mathbf{w}_\text{rad}\Bigg).
\end{align}

In terms of the interference, we consider that the approximation $\log_2(1+x) \approx x$ holds true. Following that, we approximate the expression related to $R_\text{com}$ in the following form:
\begin{equation}\label{eq:com_rate_approx}
    R_\text{com} \approx \frac{\mathbf{w}_\text{com}^H \mathbf{H}_\text{com} \mathbf{F}_\textrm{RF}^\text{com} \mathbf{S}^{\text{com}} \mathbf{F}_\textrm{BB}^\text{com} {\mathbf{F}_\textrm{BB}^\text{com}}^H \mathbf{S}^{\text{com}} {\mathbf{F}_\textrm{RF}^\text{com}}^H \mathbf{H}_\text{com}^H \mathbf{w}_\text{com}}{\sigma_\textrm{rad-com}^2},
\end{equation}
where the numerator can be written as:
\begin{align}
    \delta_\text{com} &\triangleq \mathbf{w}_\text{com}^H \mathbf{H}_\text{com} \mathbf{F}_\textrm{RF}^\text{com} \mathbf{S}^{\text{com}} \mathbf{F}_\textrm{BB}^\text{com} {\mathbf{F}_\textrm{BB}^\text{com}}^H \mathbf{S}^{\text{com}} {\mathbf{F}_\textrm{RF}^\text{com}}^H \mathbf{H}_\text{com}^H \mathbf{w}_\text{com} \nonumber \\
    &= \mathbf{w}_\text{com}^H \mathbf{H}_\text{com} \mathbf{F}_\textrm{RF}^\text{com} \mathbf{S} {\mathbf{F}_\textrm{RF}^\text{com}}^H \mathbf{H}_\text{com}^H \mathbf{w}_\text{com},\label{eq:delta_com}
\end{align}
given that $\mathbf{F}_\textrm{BB}^\text{com} {\mathbf{F}_\textrm{BB}^\text{com}}^H = \mathbf{I}$; $\mathbf{S}^{\text{com}}$ represents an idempotent matrix, i.e., $\mathbf{S}^{\text{com}} \mathbf{S}^{\text{com}} = \mathbf{S}^{\text{com}}$. 

The denominator $\sigma_\textrm{rad-com}^2$ is given by:
\begin{align}
\sigma_\textrm{rad-com}^2= \mathbf{w}_\text{rad}^H\mathbf{H}_\text{rad}\mathbf{F}_\textrm{RF}^\text{com} \mathbf{S}^{\text{com}}{\mathbf{F}_\textrm{RF}^\text{com}}^H \mathbf{H}_\text{rad}^H\mathbf{w}_\text{rad}, \label{eq:sigma_rad_com}
\end{align}
given that $\mathbb{E}\{\mathbf{s}_\text{com} \mathbf{s}_\text{com}^H \} = \mathbf{I}$. 

In similar manner for radar sensing, radar rate can be expressed as follows:
\begin{equation}\label{eq:rad_rate_approx}
    R_\text{rad} \approx \frac{\mathbf{w}_\text{rad}^H \mathbf{H}_\text{rad} \mathbf{F}_\textrm{RF}^\text{rad} \mathbf{S} \mathbf{F}_\textrm{BB}^\text{rad} {\mathbf{F}_\textrm{BB}^\text{rad}}^H \mathbf{S} {\mathbf{F}_\textrm{RF}^\text{rad}}^H \mathbf{H}_\text{rad}^H \mathbf{w}_\text{rad}}{\sigma_\textrm{com-rad}^2},
\end{equation}
where the numerator can be written as:
\begin{align}
    \delta_\text{rad} &\triangleq \mathbf{w}_\text{rad}^H \mathbf{H}_\text{rad} \mathbf{F}_\textrm{RF}^\text{rad} \mathbf{S} \mathbf{F}_\textrm{BB}^\text{rad} {\mathbf{F}_\textrm{BB}^\text{rad}}^H \mathbf{S} {\mathbf{F}_\textrm{RF}^\text{rad}}^H \mathbf{H}_\text{rad}^H \mathbf{w}_\text{rad} \nonumber\\
    &= \mathbf{w}_\text{rad}^H \mathbf{H}_\text{rad} \mathbf{F}_\textrm{RF}^\text{rad} \mathbf{S} {\mathbf{F}_\textrm{RF}^\text{rad}}^H \mathbf{H}_\text{rad}^H \mathbf{w}_\text{rad}, \label{eq:delta_rad}
\end{align}
given that $\mathbf{F}_\textrm{BB}^\text{rad} {\mathbf{F}_\textrm{BB}^\text{rad}}^H = \mathbf{I}$. Similarly for a given $\mathbb{E}\{\mathbf{s}_\text{rad} \mathbf{s}_\text{rad}^H \} = \mathbf{I}$, following above, the denominator term $\sigma_\textrm{com-rad}^2$ of the radar rate expression can be written in the following form:
\begin{align}
\sigma_\textrm{com-rad}^2 =\mathbf{w}_\text{com}^H\mathbf{H}_\text{com}\mathbf{F}_\textrm{RF}^\text{rad}\mathbf{S}^{\text{rad}}{\mathbf{F}_\textrm{RF}^\text{rad}}^H \mathbf{H}_\text{com}^H\mathbf{w}_\text{com}, \label{eq:sigma_com_rad}
\end{align}
Next, we describe our fractional programming based proposed method to optimally select the RF chains and maximize SE while minimizing the interference of one operation to the other.

\subsection{Fractional Programming}
Following \eqref{eq:com_rate_approx} and \eqref{eq:rad_rate_approx}, it can be inferred that the approximated cost functions are in fractional form. The weighted joint SE can then be approximated as:
\begin{align}
    R &\approx \rho \frac{\delta_{\text{com}}(\mathbf{S}^{\text{com}})}{\sigma_{\text{rad-com}}(\mathbf{S}^{\text{com}})} + (1-\rho) \frac{\delta_{\text{rad}}(\mathbf{S}^{\text{rad}})}{\sigma_{\text{com-rad}}(\mathbf{S}^{\text{rad}})}.
\end{align}
In order to deal with a fractional cost function, we implement the solution based on Dinkelbach (DB) iterations \cite{dinkelbach1967}, where each iteration of the DB method solves the following problem:
\begin{align}
    \max_{\mathbf{S}^{\text{com}}, \mathbf{S}^{\text{rad}}} & \left\{ \rho \delta_\text{com}(\mathbf{S}_{(i)}^{\text{com}}) + (1-\rho) \delta_\text{rad}(\mathbf{S}_{(i)}^{\text{rad}}) \right. \nonumber \\ &- \left. \kappa_{(i)}^{\text{com}} \rho \sigma_\textrm{rad-com}(\mathbf{S}_{(i)}^{\text{com}}) - \kappa_{(i)}^{\text{rad}} (1-\rho) \sigma_\textrm{com-rad}(\mathbf{S}_{(i)}^{\text{rad}})\right\} \nonumber \\
    \textrm{ subject to } &[\mathbf{S}_{(i)}^{\text{com}}]_{k,k}, [\mathbf{S}_{(i)}^{\text{rad}}]_{k,k} \in [0,1], \label{eq:S_problem_com}
\end{align}
for $i=1,2,\ldots,I_{\rm max}$, where
\begin{equation} \label{eq:kappa_computation}
    \kappa_{(i)}^{\text{com}} = \delta_\text{com}^{(i-1)}/\sigma_\text{rad-com}^{(i-1)}, \text{ and }     \kappa_{(i)}^{\text{rad}} = \delta_\text{rad}^{(i-1)}/\sigma_\text{com-rad}^{(i-1)},
\end{equation}
with $\mathbf{S}_{(0)}^{\text{com}}=\mathbf{S}_{(0)}^{\text{rad}}=\mathbf{I}$, $\kappa_{(0)}^{\text{com}}=\kappa_{(0)}^{\text{rad}}=1$.

The DB method has been widely implemented in fractional problems which yields good performance \cite{aryanTGCN2019, aryanTGCN2021}. The step-wise DB-based procedure of the proposed algorithm which provides solution to \eqref{eq:S_problem_com} and selects the optimal number of RF chains during communication operation with reduced radar interference is shown in Algorithm \ref{algorithm:proposed}. Next, we present the simulation results and comparison with existing baselines while taking into account different interference scenarios..

\section{Simulation Results}
This section discusses simulation results supporting the significance of our proposed flexible hybrid beamforming based RF chain selection approach. In terms of simulation parameters to observe different performance curves, MIMO JRC has been implemented where the number of transmitter antennas $N_\textrm{T} = 96$ and the number of receiver (UE) antennas $N_\textrm{R} = 4$. The number of streams $N_\textrm{s} = N_\textrm{R}$, and the number of multipaths $N_\textrm{c} = 6$. For radar sensing operation, the number of targets $N_\textrm{p} = 3$, and angular target locations are random for each realization. The total number of available RF chains is considered to be equal to $N_\textrm{R}$. The precoding matrices $\mathbf{F}_{\text{BB}}=\mathbf{I}$, and $\mathbf{F}_{\text{RF}}$ is considered to be fast Fourier transform (FFT) matrix for both radar and communications. It is assumed that the JRC system implements ULA setup at both transmitter and receiver (UE) sides. To focus mainly on performance of the transmitter side, we assume that digital combining is performed at the receiver, i.e., $\mathbf{w}_k$ is expressed as the $k^\mathrm{th}$ column of the left orthonormal matrix, which is obtained by the singular value decomposition of channel matrix. The proposed method implements hybrid precoding and RF chain selection procedure for SE maximization which takes into account both the interference from radar to communication, and vice-versa. 

To compare the proposed flexible hybrid beamforming based method and observe its effectiveness, we consider four baseline schemes which are based on hybrid precoding, however differ in terms of interference:\\ 
\emph{(i) no interference}, i.e., hybrid case with no interference from radar or communications; \\
\emph{(ii) interference both for comms and radar}, i.e., hybrid case with fixed number of available RF chains ($L_\textrm{T} = N_\textrm{R}$) and when interference from both radar and communication operations is considered; \\
\emph{(iii) interference only for radar}, i.e., hybrid case when there is communication interference to the radar operation; \\
\emph{(iv) interference only for comms}, i.e., hybrid case when there is radar interference to the communication operation.

\begin{figure}[t]
    \centering
    \includegraphics[scale=0.285]{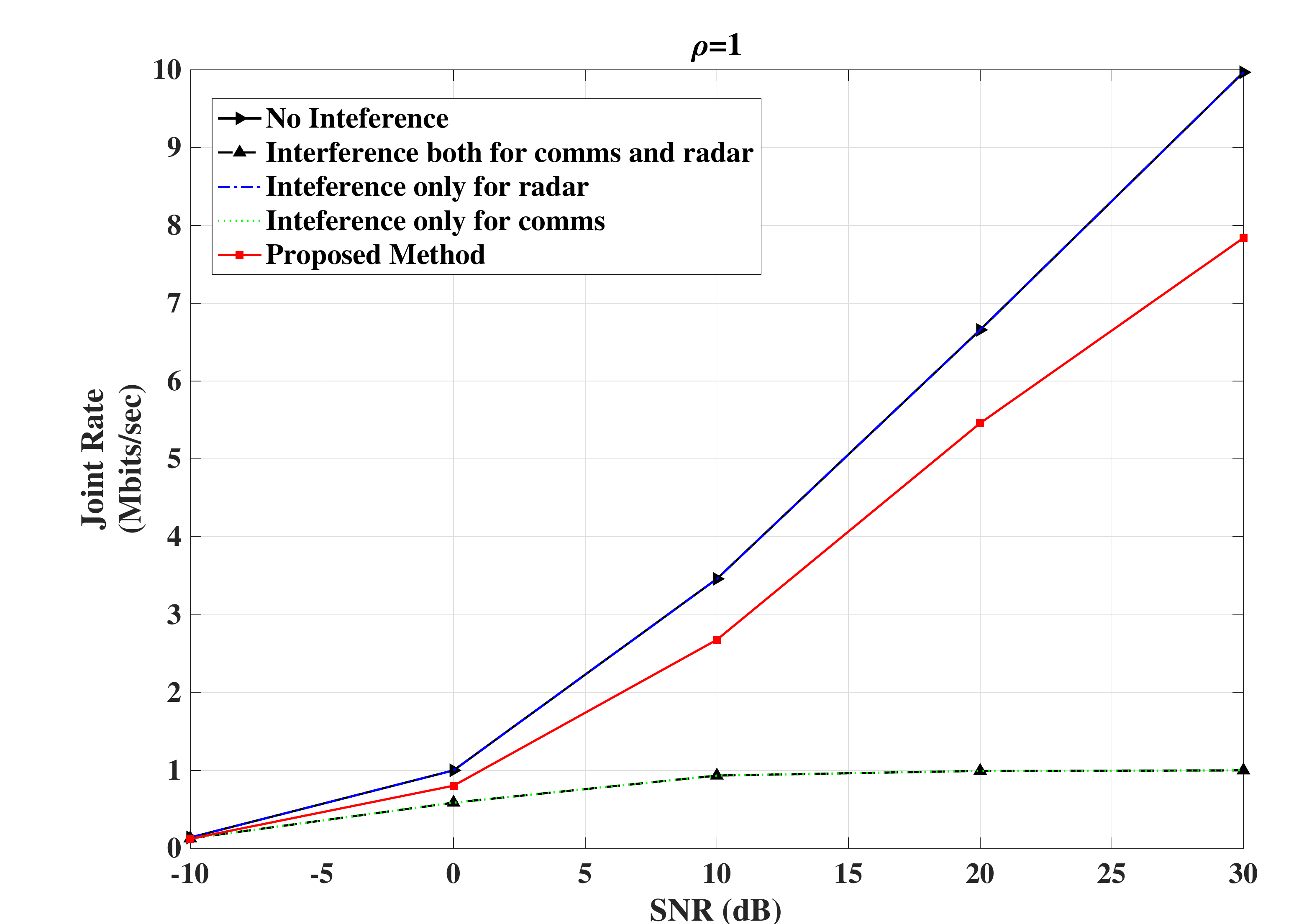}
    \caption{Joint rate w.r.t. SNR for $N_\textrm{T} = 96$, $N_\textrm{R} = 4$ and weighting factor $\rho = 1$.}
    \label{fig:1}
\end{figure}

\begin{figure}[t]
    \centering
    \includegraphics[scale=0.285]{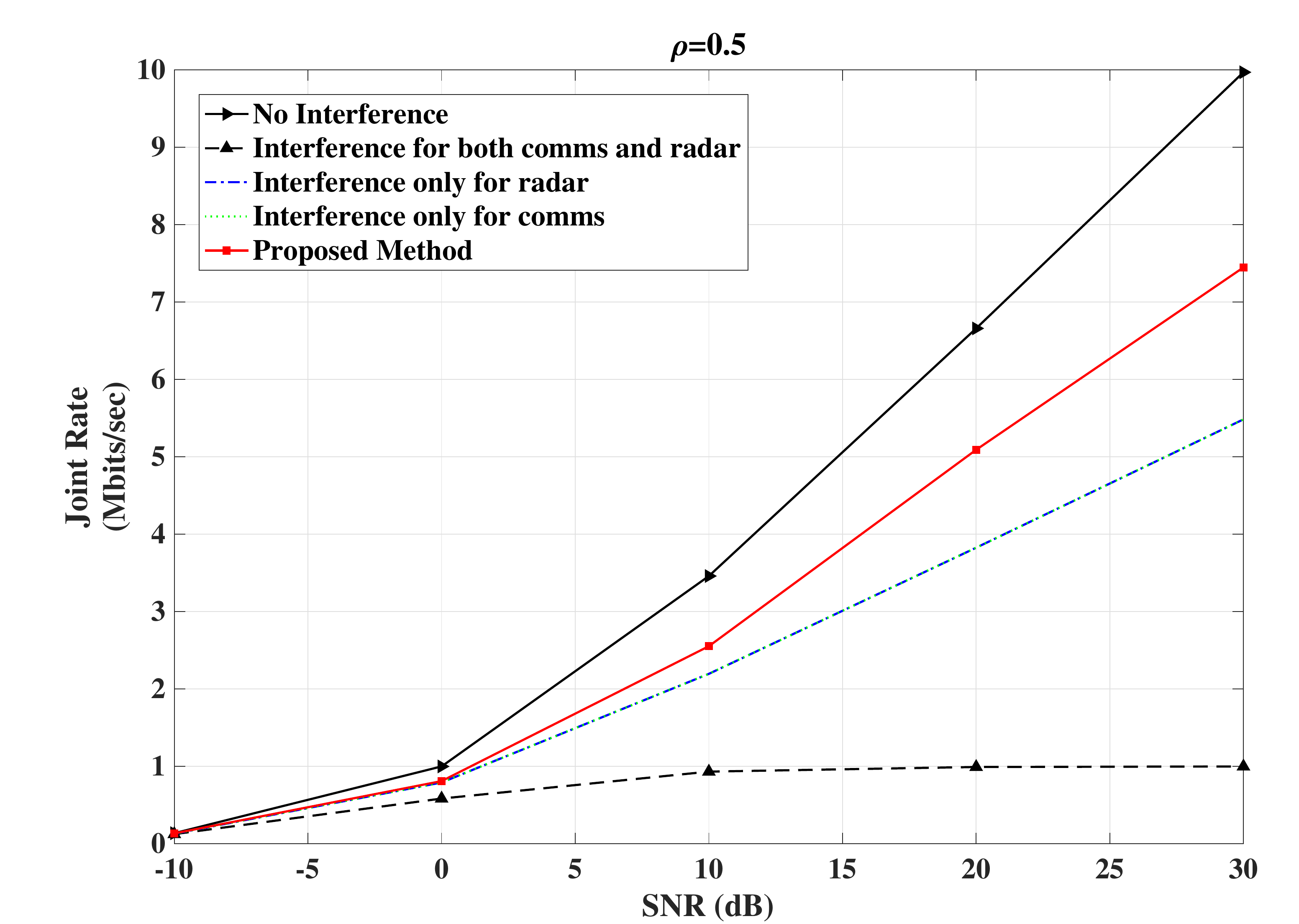}
     \caption{Joint rate w.r.t. SNR for $N_\textrm{T} = 96$, $N_\textrm{R} = 4$ and weighting factor $\rho = 0.5$.}
    \label{fig:2}
\end{figure}

\begin{figure}[t]
    \centering
    \includegraphics[scale=0.285]{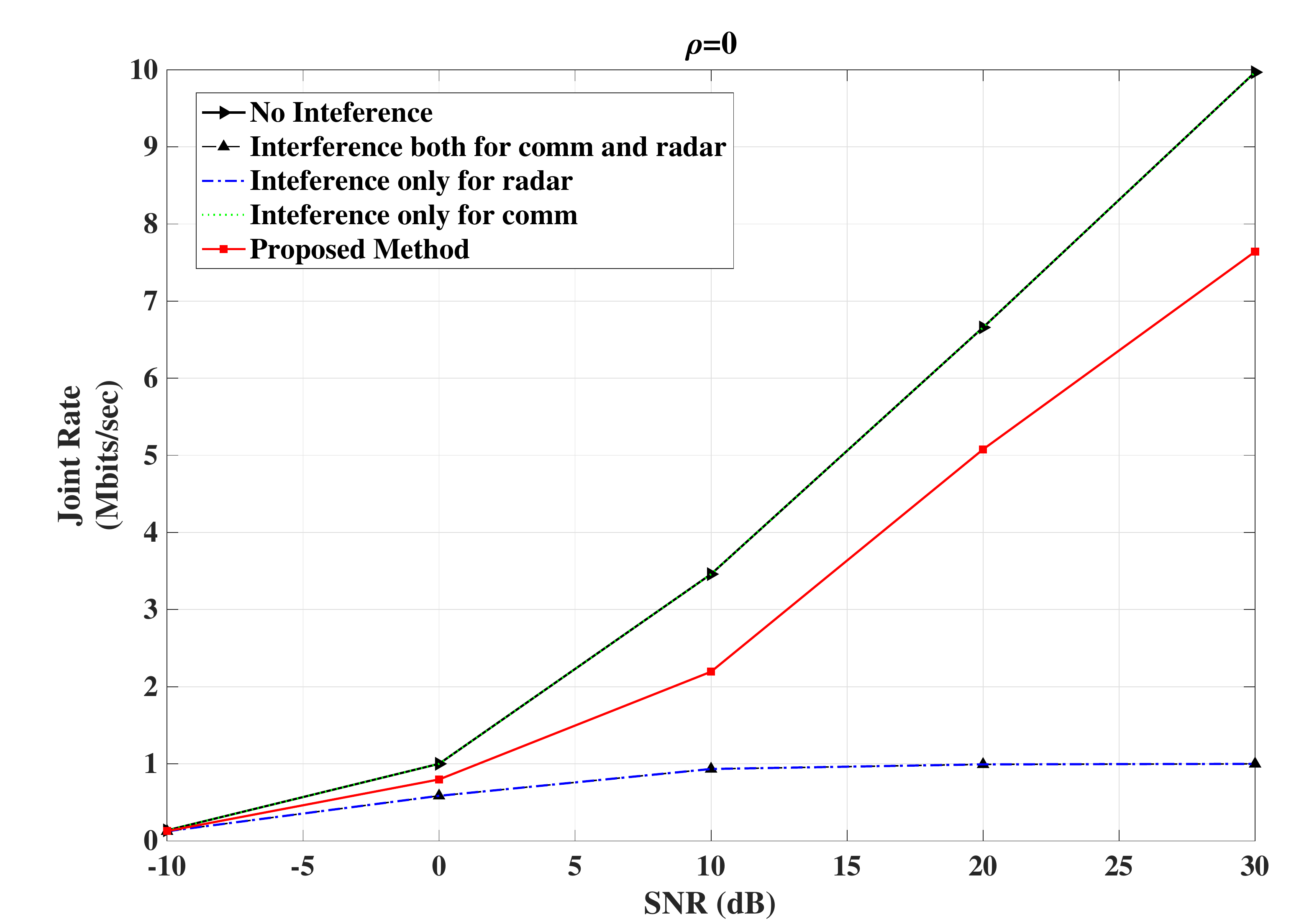}
    \caption{Joint rate w.r.t. SNR for $N_\textrm{T} = 96$, $N_\textrm{R} = 4$ and weighting factor $\rho = 0$.}
    \label{fig:3}
\end{figure}

\begin{figure}[t]
    \centering
    \includegraphics[scale=0.29]{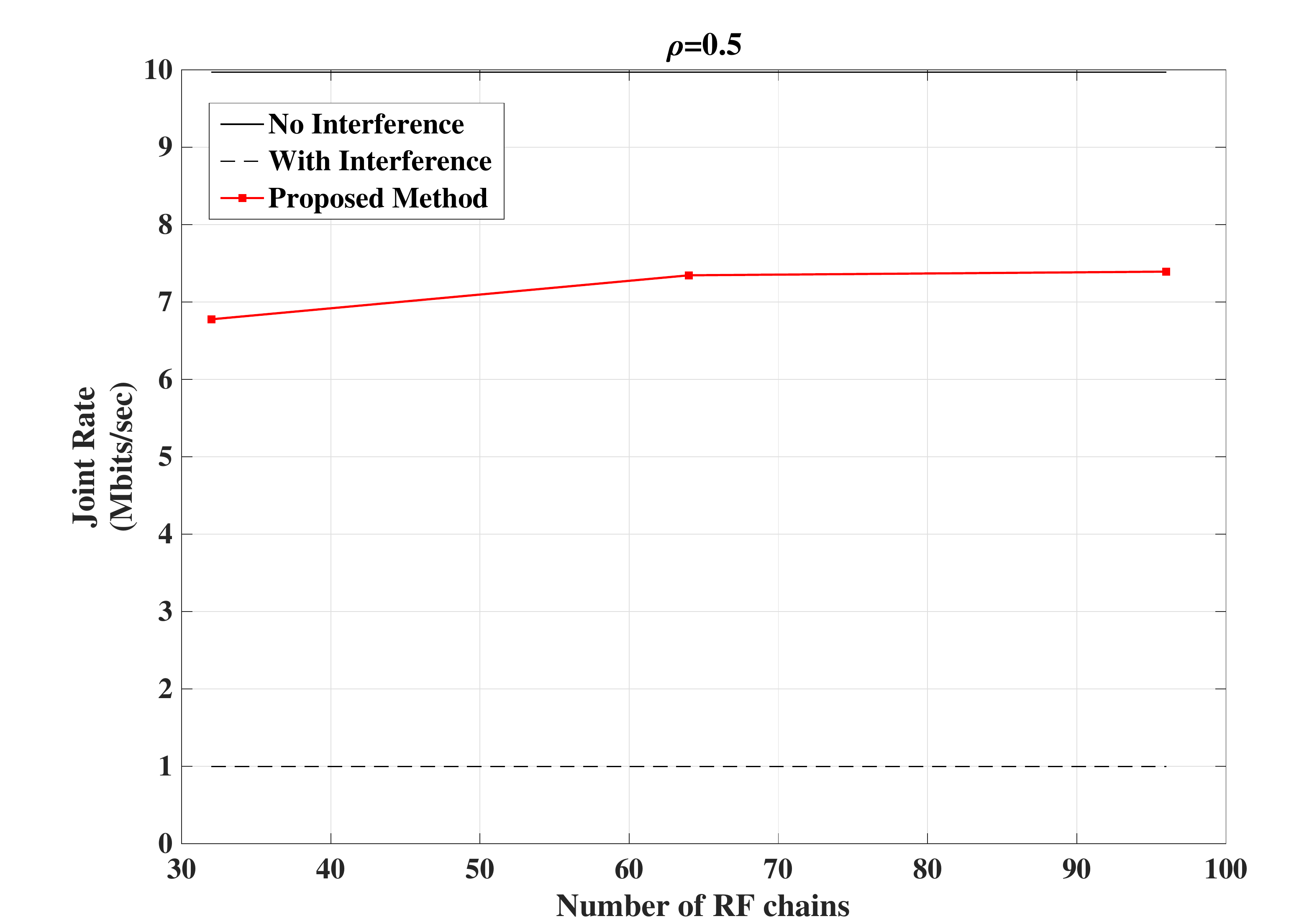}
     \caption{Joint rate w.r.t. number of RF chains $N_\textrm{T} = 96$, $N_\textrm{R} = 4$ and weighting factor $\rho = 0.5$.}
    \label{fig:4}
\end{figure}

From Fig. 2, we can observe the joint rate performance with respect to (w.r.t.) signal-to-noise ratio (SNR) at $N_\textrm{T} = 96$, $N_\textrm{R} = 4$, for the proposed method and compare it with the above mentioned baseline cases. It can be observed that the proposed method performs better than interference-oriented baseline cases such as ii) and iv) above, and approximates the rate performance when there is no interference case, specially for low-SNR region values. For instance, at SNR = 0 dB, it can be observed that the proposed method approximates baselines (i) and (iii) with joint rate being nearly 1 Mbits/sec. Note that the weighting factor considered in Fig. 2 is $\rho = 1$ which means communication-only operation takes place and there is no radar operation that takes place.

Similarly for Fig. 3, joint rate performance w.r.t. SNR is observed at $N_\textrm{T} = 96$, $N_\textrm{R} = 4$ and weighting factor $\rho = 0.5$, which means the radar and communication operations have the same priority. Similar performance pattern to Fig. 2 can be observed in terms of joint rate for the proposed method which approximates the baseline case (i) with no interference and better than other baselines. Similarly for Fig. 4, when weighting factor is changed to $\rho = 0$, radar-only operation takes place, i.e., there is no communication operation, and good performance is exhibited by the proposed method. Fig. 5 plots the joint rate performance w.r.t. the number of RF chains for a weighting factor $\rho=0.5$ (equal priority to both the operations), $N_\textrm{T} = 96$ and $N_\textrm{R} = 4$. The rate variation for the proposed method can be observed which is also compared with both interference (which includes both radar and communication interference) and no interference cases.

\section{Conclusion}
This paper designs joint SE maximization problem for hybrid precoding based MIMO JRC systems with dual function radar and communication operations. We consider interference from one operation to the other in our problem formulation, and implement an optimal RF chain selection procedure for flexible hybrid beamforming design. The proposed method based on fractional programming yields good joint rate performance when compared with existing baselines which implement fixed number of RF chains and different interference cases. The effect of weighting factor which prioritizes one operation over the other, has also been observed via numerical results and comparison with baselines.



\bibliographystyle{IEEEtran}

\end{document}